\documentclass[a4paper]{jpconf}
\usepackage{graphicx}
\begin{document}
\title{Neutron diffraction in a model itinerant metal near a quantum critical point}

\author{D A Sokolov$^1$, M C Aronson$^1$$^,$$^2$, R Erwin$^3$, J W Lynn$^3$, M D Lumsden$^4$ and S E Nagler$^4$}

\address{$^1$ Brookhaven National Laboratory, Upton, NY 11973, USA}
\address{$^2$ Department of Physics and Astronomy, Stony Brook University, Stony Brook, NY 11794, USA}
\address{$^3$ NIST Center for Neutron Research, Gaithersburg, MD 20899, USA}
\address{$^4$ Neutron Scattering Science Division, Oak Ridge, TN 37831, USA}

\ead{fermiliquid@gmail.com}

\begin{abstract}
Neutron diffraction measurements on single crystals of
Cr$_{1-x}$V$_{x}$ (x=0, 0.02, 0.037) show that the ordering moment
and the Neel temperature are continuously suppressed as x approaches
0.037, a proposed Quantum Critical Point (QCP). The wave vector Q of
the spin density wave (SDW) becomes more incommensurate as x
increases in accordance with the two band model. At x$_{C}$=0.037 we
have found temperature dependent, resolution limited elastic
scattering at 4 incommensurate wave vectors Q=(1$\pm\delta$$_{1,2}$,
0, 0)*2$\pi$/a, which correspond to 2 SDWs with Neel temperatures of
19 K and 300 K. Our neutron diffraction measurements indicate that
the electronic structure of Cr is robust, and that tuning Cr to its
QCP results not in the suppression of antiferromagnetism, but
instead enables new spin ordering due to novel nesting of the Fermi
surface of Cr.
\end{abstract}

\section{Introduction}
Cr doped with V has been proposed as an example of an itinerant
antiferromagnet, with a second order quantum phase
transition~\cite{<1>}. A gradual evolution of the Hall coefficient
across the Quantum Critical Point (QCP) was observed, leading to the
conclusion that the spin density wave order in Cr$_{1-x}$V$_{x}$
develops continuously~\cite{<2>}. Spin density order in Cr removes
the hot spots of the Fermi surface, which leads to a restructured or
nested Fermi surface. In a neutron diffraction experiment, magnetic
Bragg scattering is detected at wave vectors equal to nesting wave
vectors of the Fermi surface, and its temperature dependence, as
well as that of the ordered moment are both measures of the Fermi
surface removed by the spin density wave. Although Cr$_{1-x}$V$_{x}$
was proposed as a model system in which the quantum phase transition
of an itinerant antiferromagnet is realized, a detailed neutron
scattering study of the spin density wave instability near the
proposed quantum critical point in Cr$_{1-x}$V$_{x}$ is still
lacking. Here we report direct measurements of spin density wave
ordering by neutron diffraction, carried out on single crystals of
Cr$_{1-x}$V$_{x}$. Our results support the general notion that a
quantum phase transition is accompanied by the restructuring of the
Fermi surface~\cite{<3>}. The new nesting conditions enabled by the
V doping in Cr are responsible for a topological change of the Fermi
surface across the QCP.

\section{Experimental Details}
Single crystals of Cr$_{1-x}$V$_{x}$ (x=0.0, 0.02, 0.037), were
grown by the arc zone melting method at the Materials Preparation
Center at Ames National Lab. The uniformity of the V doping in the
crystals of Cr$_{1-x}$V$_{x}$ was confirmed by electron microprobe
and electron energy loss  measurements. Neutron diffraction
measurements were carried out on single crystals of
Cr$_{1-x}$V$_{x}$ (x=0.0, 0.02, 0.037) using the BT-9 triple-axis
spectrometer at the NIST Center for Neutron Research with a fixed
final energy E$_{F}$=14.7 meV. The measurements were performed using
a 40'-44'-44'-open collimation configuration. Data were collected
near the (1,0,0) and (0,1,0) reciprocal lattice positions in the
(001) plane.

\section{Results and Discussion}
Pure Cr orders antiferromagnetically at 311 K via a spin density
wave (SDW)instability. A wavevector $\emph{Q}$ for the SDW is
selected by the nesting condition of the Fermi surface, which
consists of electron and hole octahedra. Since the hole Fermi
surface is slightly larger than the electron Fermi surface,
$\emph{Q}$ is incommensurate with the lattice. In a neutron
diffraction experiment, the incommensurability of $\emph{Q}$ leads
to an observation of 2 magnetic Bragg reflections at
q=2$\pi$/a(1$\pm\delta$,0,0). In the absence of a preferred
orientation, $\emph{Q}$ can lie along any of the 3 cubic axis so in
total, 6 satellites can be found near q=2$\pi$/a(1,0,0).

Fig.~1a shows the incommensurate satellites near q=2$\pi$/a(1,0,0)
and q=2$\pi$/a(0,1,0), which were probed by transverse and
longitudinal scans in pure Cr and Cr$_{0.98}$V$_{0.02}$ below their
respective ordering temperatures. Fig.~1b shows a reciprocal plane
for Cr$_{0.963}$V$_{0.037}$. At 5 K a longitudinal scan in the
$\emph{h00}$ direction finds two satellites at
q$_{1}$=2$\pi$/a(1$\pm\delta_{1}$,0,0) and two at
q$_{2}$=2$\pi$/a(1$\pm\delta_{2}$,0,0). A transverse scan in the
$\emph{1k0}$ direction finds two satellites at
q$_{2}$=2$\pi$/a(1,0$\pm\delta_{2}$,0) only. The elastic scattering
in Cr$_{0.963}$V$_{0.037}$ is resolution limited with respect to the
wave vector, corresponding to antiferromagnetic order extending over
length scales as large as $\sim$34$\AA$, unexpected for a system
very near a QCP.

\begin{figure}[h]
\includegraphics[width=22pc]{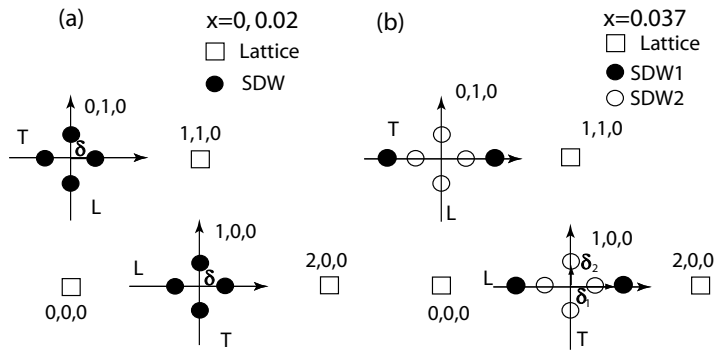}\hspace{2pc}%
\begin{minipage}[b]{12pc}\caption{\label{label}The 001 reciprocal planes of
Cr$_{1-x}$V$_{x}$ probed in our experiments. (a) x=0, 0.02, L and T
mark the longitudinal and transverse directions of scans. $\delta$
is the incommensurability parameter. (b) Same for x=0.037.}
\end{minipage}
\end{figure}

We refer to scattering at q$_{1}$=2$\pi$/a(1$\pm\delta_{1}$,0,0) as
originating with SDW1, while that at
q$_{2}$=2$\pi$/a(1$\pm\delta_{2}$,0,0) originates with SDW2. Since
the scattering at q$_{1}$=2$\pi$/a(1$\pm\delta_{1}$,0,0) is only
observed on longitudinal scans through (1,0,0) or transverse scans
through (0,1,0) points in the reciprocal lattice, SDW1 must be
transversely polarized with a magnetic moment along (0,0,1). The
intensities of the incommensurate satellites at
q$_{1}$=2$\pi$/a(1$\pm\delta_{1}$,0,0) obey the magnetic form
factor, confirming that the scattering is magnetic.

In Fig.~2a we plot the temperature dependence of the intensity of
the incommensurate satellites in Cr$_{1-x}$V$_{x}$ (x=0.0, 0.02,
0.037). In pure Cr the transition into the SDW state is weakly first
order,  therefore the incommensurate scattering increases sharply
below T$_{N}$=311 K. Below 122 K, the SDW changes polarisation from
transverse to longitudinal, which leads to a loss of intensity at
q=2$\pi$/a(1$\pm\delta$,0,0). In Cr$_{0.98}$V$_{0.02}$ the
incommensurate intensity develops gradually below T$_{N}$=145 K as
an order parameter, indicating that V doping renders the
antiferromagnetic transition second order. The intensity increases
monotonically on cooling to 5 K suggesting that the polarisation of
the SDW in Cr$_{0.98}$V$_{0.02}$ is transverse between 5 K and 145
K.

The incommensurate scattering in Cr$_{0.963}$V$_{0.037}$ is more
complex. Scattering at q$_{1}$=2$\pi$/a(1$\pm\delta_{1}$,0,0) which
corresponds to the most incommensurate satellites, turns on sharply
at 19 K, a temperature consistent with the transition temperature
obtained from the earlier resistivity measurements~\cite{<4>}. In
contrast, the temperature dependence of the incommensurate
scattering at q$_{2}$=2$\pi$/a(1$\pm\delta_{2}$,0,0) is that of an
order parameter with a transition temperature of $\sim$ 300 K. The
intensities of scattering at q=2$\pi$/a(1$\pm\delta_{2}$,0,0) and at
q=2$\pi$/a(0,1$\pm\delta_{2}$,0) are very similar indicating that
our sample is in a multiple-\emph{Q} state. Below T=19 K, SDW1 and
SDW2 co-exist.

\begin{figure}[h]
\includegraphics[width=26pc]{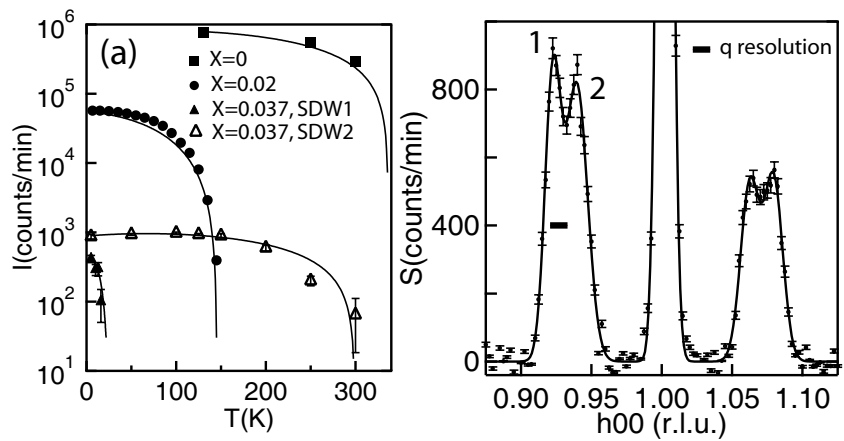}\hspace{2pc}%
\begin{minipage}[b]{12pc}\caption{\label{label}(a) Temperature dependence
of the incommensurate scattering in Cr$_{1-x}$V$_{x}$ x=0.0, 0.02,
0.037. Solid lines are order parameter fits. (b) Elastic
longitudinal scan through (1,0,0) in Cr$_{0.963}$V$_{0.037}$
collected at 5 K.  Bragg peaks 1 and 2 mark SDW1 and SDW2. The solid
line is a fit to the data where the fitting function is a sum of 5
gaussian functions plus a linear background. The horizontal solid
line is the wave vector resolution of BT-9.}
\end{minipage}
\end{figure}

A longitudinal scan through the (1,0,0) point in the reciprocal
lattice collected at 5 K in Cr$_{0.963}$V$_{0.037}$ is shown in
Fig.~2b. Two separate incommensurate peaks corresponding to SDW1 and
SDW2 can be distinguished. The strong commensurate scattering is due
to temperature independent $\lambda$/2 contamination from the (200)
structural reflection as well as temperature dependent contamination
from the q=2$\pi$/a(1,0,0$\pm\delta$) satellites, which can be
suppressed by tightening of the vertical resolution of the
spectrometer.

\begin{figure}[h]
\includegraphics[width=14pc]{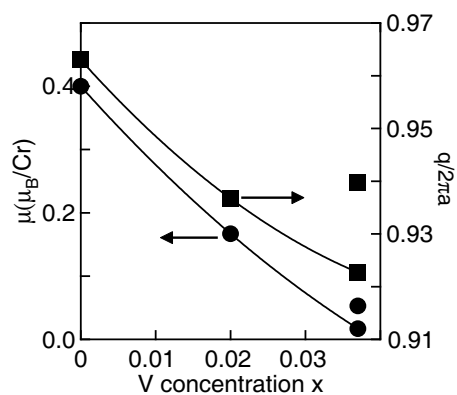}\hspace{2pc}%
\begin{minipage}[b]{12pc}\caption{\label{label}
The wavevector of the SDW modulation and the ordering moment in
Cr$_{1-x}$V$_{x}$ as a function of the V concentration measured at 5
K. Solid lines are guides to the eye. Error bars are the size of a
marker, or otherwise shown.}
\end{minipage}
\end{figure}

To estimate the ordered moment in Cr$_{0.963}$V$_{0.037}$, we have
carried out identical neutron diffraction measurements on a  single
crystal of pure Cr of similar size and shape as the x=0.037 doped
crystals. The intensities of the (2,0,0) and (1,1,0) structural
peaks were very similar in the two crystals, allowing a direct
comparison between integrated intensities of the q=2$\pi$/a(1,0,0)
peaks in Cr and Cr$_{0.963}$V$_{0.037}$. The same procedure was
followed to estimate the ordered moment in Cr$_{0.98}$V$_{0.02}$.
The average moment per Cr in SDW1 in Cr$_{0.963}$V$_{0.037}$
$\langle\mu\rangle$=0.017 $\mu_{B}$/Cr at 5 K, a much smaller value
than the $\langle\mu\rangle$=0.4 $\mu_{B}$/Cr found in pure
Cr~\cite{<5>}. In Fig.~3 we show that as the V concentration
increases, the ordered moment of the SDW is suppressed and the
wavevector becomes more incommensurate. Instead of continuously
approaching zero at the QCP, we find that as x$\rightarrow$0.04
Cr$_{1-x}$V$_{x}$ develops 2 SDW instabilities at 2 different
wavevectors.

It is instructive to compare the neutron diffraction data on V doped
Cr with the results of X-ray scattering studies on pure Cr under
pressure~\cite{<6>}, Fig.~4. Since V has one less electron than Cr,
V doping can broadly be considered to expand the hole Fermi surface,
driving the ordering wave vector more incommensurate and by reducing
the fraction of the total Fermi surface which is nestable,
destabilizes antiferromagnetic order and reduces T$_{N}$. In
contrast, pressure is not thought to significantly change the
relative sizes of the electron and hole Fermi surfaces, so the
nesting and the Neel temperature are expected to be less strongly
affected. Both pressure and doping decrease the lattice parameters,
so it is possible to directly compare their respective impacts on
the magnetic order at identical lattice spacings. We have compiled
values of the Neel temperature in V-doped samples from both our own
samples and from the literature, and have compared them to the
values derived from high pressure x-ray diffraction measurements in
Fig.~4. As the standard electronic model predicts, V doping
suppresses T$_{N}$ much more rapidly than pressure. Whereas the
applied pressure monotonically reduces the T$_{N}$ from 311 K to
$\sim$90 K at 7 GPa, the V doping reduces the T$_{N}$ to 19 K at a
doping level of only 3.75$\%$.  However, the standard electronic
model, where nesting of the electron and hole Fermi surfaces
stabilizes a single SDW, we find a very different situation near the
doping induced quantum critical point. Here, the conventional SDW
removes so little of the Fermi surface that the energetics of Cr-V
enable a new Fermi surface driven instability, occuring at high
temperatures and at a different wave vector, perhaps involving
nesting which spans the individual electron or hole surfaces.

\begin{figure}[h]
\includegraphics[width=14pc]{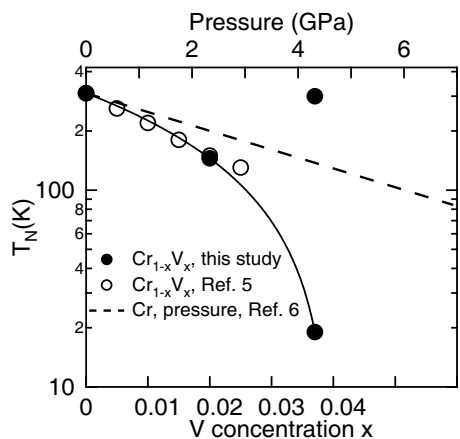}\hspace{2pc}%
\begin{minipage}[b]{12pc}\caption{\label{label}
Effects of V doping and external pressure on T$_{N}$ in Cr. Dashed
line represents the data from~\cite{<6>}. Solid lines are guides to
the eye. Error bars are the size of a marker, otherwise shown.}
\end{minipage}
\end{figure}

In summary, we have shown through neutron diffraction measurements
of the spin density wave instabilities that antiferromagnetic order
does not vanish near the quantum critical point in Cr which was
proposed at the doping level x$\leq$3.75$\%$. Instead we find that
Cr$_{0.963}$V$_{0.037}$ is a highly ordered system with a new type
of antiferromagnetic ordering which persists up to 300 K. This means
that the simple Fermi surface nesting model used to describe the SDW
instability is no longer appropriate near the quantum critical point
and that new theoretical models are needed to explain these novel
and co-existing ground states.

\subsection{Acknowledgments}
D. A. S. and M. C. A. would like to thank A. M. Tsvelik, T. M. Rice,
and S. M. Shapiro for useful discussions. Work at BNL and ORNL is
supported by the Department of Energy. Work at Stony Brook
University is supported by the National Science Foundation.


\section*{References}
\begin{thebibliography}{<10>}
\bibitem{<1>}Yeh A, Soh Y, Brooke J, Aeppli G, Rosenbaum T F and Hayden S M 2002 {\it Nature} $\bf{419}$ 459
\bibitem{<2>}Lee M, Husmann A, Rosenbaum T F and Aeppli G 2004 {\it Phys. Rev. Lett.} $\bf{92}$ 187201
\bibitem{<3>}Gegenwart P, Si Q and Steglich F 2008 {\it Nature Physics} {\bf 4} 186
\bibitem{<4>}Sokolov D A, Aronson M C, Strycker G L, Lumsden M D, Nagler S E and Erwin R 2008 {\it Physica} B {\bf 403} 1276
\bibitem{<5>}Koehler W C, Moon R M, Trego A L and Mackintosh A R 1966 {\it Phys. Rev.} $\bf{151}$ 405.
\bibitem{<6>}Feng Y, Jaramillo R, Srajer G, Lang J C, Islam Z, Somayazulu M S, Shpyrko O G, Pluth J J, Mao H -k, Isaacs E D, Aeppli G and Rosenbaum T F 2007 {\it Phys. Rev. Lett.} $\bf{99}$ 137201
\end{thebibliography}
\end{document}